**Asymmetric Schottky Contacts in Bilayer MoS$_2$ Field Effect Transistors**

*Antonio Di Bartolomeo\*, Alessandro Grillo, Francesca Urban, Laura Iemmo, Filippo Giubileo, Giuseppe Luongo, Giampiero Amato, Luca Croin, Linfeng Sun, Shi-Jun Liang, and Lay Kee Ang*

Prof. A. Di Bartolomeo, A. Grillo, F. Urban, Dr. L. Iemmo, G. Luongo

Physics Department, University of Salerno, and CNR-SPIN, Fisciano 84084, Italy

E-mail: adibartolomeo@unisa.it

Prof. F. Giubileo

CNR-SPIN Salerno, via Giovanni Paolo II n. 132, Fisciano 84084, Italy

Prof. G. Amato, Dr. L. Croin

Istituto Nazionale di Ricerca Metrologica, INRIM - Strada delle Cacce, Torino 10135, Italy

Dr. L. Sun

Department of Energy Science, Sungkyunkwan University, Suwon 16419, Korea

Dr. S.-J. Liang

Engineering Product Development (EPD), Singapore University of Technology and Design (SUTD), 487372 Singapore

National Laboratory of Solid State Microstructures, School of Physics, Collaborative Innovation Center of Advanced Microstructures, Nanjing University, Nanjing 210093, China

Prof. L. K. Ang

Engineering Product Development (EPD), Singapore University of Technology and Design (SUTD), 487372 Singapore

Abstract: We discuss the high-bias electrical characteristics of back-gated field-effect transistors with CVD-synthesized bilayer $MoS_2$ channel and Ti Schottky contacts. We find that oxidized Ti contacts on $MoS_2$ form rectifying junctions with ~0.3 to 0.5 eV Schottky barrier height. To explain the rectifying output characteristics of the transistors, we propose a model based on two slightly asymmetric back-to-back Schottky barriers, where the highest current arises from image force barrier lowering at the electrically forced junction, while the reverse current is due to Schottky-barrier limited injection at the grounded junction. The device achieves a photo responsivity greater than 2.5 $AW^{-1}$ under 5 $mWcm^{-2}$ white-LED light. By comparing two- and four-probe measurements, we demonstrate that the hysteresis and persistent photoconductivity exhibited by the transistor are peculiarities of the $MoS_2$ channel rather than effects of the $Ti/MoS_2$ interface.

## 1. Introduction

Molybdenum disulfide ($MoS_2$), a member of the transition metal dichalcogenides family, has recently become one of the most promising layered materials for next generation of electronic devices and sensors as alternative or complement to graphene. Similarly to graphene,[1-5] $MoS_2$- based transistors,[6] integrated circuits,[7,8] photodetectors,[9-11] solar cells,[12] supercapacitors,[13] and chemical[14] or biological[15] sensors have been reported.

Monolayer or few-layers $MoS_2$ can be mechanically exfoliated and transferred on a substrate for device fabrication; however, large-scale, high-crystalline and uniform monolayer or few layers $MoS_2$ flakes, up to the centimeter size, are nowadays produced by chemical vapor deposition (CVD) from solid state precursors.[16]

$MoS_2$ has a layered structure where monolayers, constituted by a plane of Mo atoms sandwiched between two planes of S atoms, are piled on top of each other bonded by weak van der Waals interaction. [17] In its bulk form, $MoS_2$ is an indirect semiconductor with bandgap of 1.2-1.3 eV, while monolayer $MoS_2$ becomes a direct semiconductor with bandgap increasing up to 1.8-1.9 eV.[18,19] The high bandgap constitutes a noteworthy advantage of this material, enabling for instance high performance switching when used as the channel in field effect transistors (FETs) or selective light absorption in optoelectronic applications. To date, broad applications of $MoS_2$ have been hindered by undesirable features, such as the low carrier mobility in the order of few hundreds $cm^2V^{-1}s^{-1}$,[20] and the sensitivity to oxygen, water or other adsorbates that make $MoS_2$ devices unstable. [21,22]

Bilayer $MoS_2$ has a direct bandgap of 1.6-1.7 eV,[23] which is electrically tunable.[24] Although it seems to have lower mobility than its monolayer counterpart, bilayer MoS2 is easier to obtain by CVD, is less affected by ambient exposure,[25] is more robust and tends to form smaller Schottky barriers with contact metals, due to the interlayer coupling, which enables higher electron injection efficiency.[26] In this regard, bilayer MoS2 offers a better trade-off between the reduced off-state current and the on-state current in transistors for integrated logic circuits.[27]

The formation of Schottky contacts of mono- or bilayer $MoS_2$ with most metals is favored by the high bandgap.[28] The Schottky barrier is usually caused by adsorbates at the interface. Both Schottky contacts and air exposure are responsible for mobility degradation and increase in the contact resistance, which are detrimental for the realization of high performance devices. Along this line, most of the recent research has been focused on the reduction of the contact resistance in ohmic contacts through suitable materials and processes,[29-31] or on the mobility enhancement by dielectric screening through $MoS_2$ channel encapsulation in protective layers.[32] Conversely, the properties of Schottky contacts on $MoS_2$ and the corresponding devices have not been fully investigated and are now attracting growing attention. Although seen as a shortcoming of the transistor performance, if well understood, Schottky contacts could enable new functionalities. For instance, Schottky barriers can be used to reduce the transistor off current, be exploited for photodetection,[33] in photovoltaic cells [34,35] or for bio-chemical sensing purposes.[36]

Schottky contacts to monolayer $MoS_2$ with ferromagnetic Co electrodes have been investigated for spintronic applications and control over the Schottky barrier height by insertion of a thin MgO oxide layer, as well as the dependence of the Schottky barrier on the applied back gate voltage, has been demonstrated.[37] N-type Schottky behavior of Au and Pd contacts has been reported both on mono and multi-layer $MoS_2$.[38] The inhomogeneity of the Au/$MoS_2$ Schottky barrier has been investigated very recently and attributed to defects, charge puddles and grain boundaries at the contact.[39] Using density functional calculations, it has been suggested that the metal-$MoS_2$ interaction makes the Fermi level pinned at 0.1-0.3 eV below the conduction band edge of $MoS_2$ for low work function metals, while for high work function metals ($\gtrsim 4.7$ eV) the Schottky barrier generally obeys the Schottky-Mott rule.[40] Similarly to MgO, it has also been demonstrated that inserting a boron nitride monolayer can prevent the interaction between metal and $MoS_2$, thereby reducing the Schottky barrier height.[41] A Schottky barrier, decreasing with the increasing number of $MoS_2$ layers, has been reported for Al contacts.[42] Gr/$MoS_2$ Schottky barriers have been also studied as rectifying heterojunction[43, 44] or transistors.[45]

In this paper, we study carrier transport through Ti Schottky contacts, due to metal oxidation, in back-gated CVD synthesized MoS$_2$-bilayer field-effect transistors. The I-V characteristic curves show a rectifying behavior, which we exploit to achieves a photo responsivity greater than 2.5 AW$^{-1}$ under 5 mWcm$^{-2}$ white-LED light. We propose a model based on two back-to-back Schottky junctions to explain the observed electrical characteristics. In particular, we unveil the importance of the image force Schottky barrier lowering at high bias in the electrical conduction through the channel of the MoS$_2$ transistor. We compare two and four-probe measurements to highlight effects such as the hysteresis and the persistent photoconductivity, which we identify as properties of the MoS$_2$ channel rather than of the rectifying contacts.

This study represents an advance in the comprehension of the transport mechanisms at the metal/MoS$_2$ Schottky contacts. It confirms the time-degradation due to ambient exposure, which has to be taken into account for any real application of MoS$_2$ in electronic or spintronic devices or in sensors. Finally, the paper highlights the potentiality of Schottky contacts on MoS$_2$ for photodetection.

## 2. Results and discussion

### 2.1. Device fabrication

MoS$_2$ flakes were synthesized by chemical vapor deposition on SiO$_2$/p-Si substrate. **Figure 1**(a) shows the scanning electron microscopy (SEM) image of a selected flake with Ti/Au metal contacts deposited by standard electron-beam lithography (EBL) and lift-off process. Unreacted residuals of MoO3 are visible on the flake as brighter dot branched sequences. The schematic cross-section of the device, consisting of a back-gated TLM (transfer length method) structure, is shown in Figure 1(b). The MoS$_2$ flake is characterized by micro-Raman. The spectrum and a map, exhibiting the $A_{1g}$ and the $E_{2g}^1$ peaks at a frequency separation of $21-22$ cm$^{-1}$, fingerprint of a bilayer MoS$_2$,[46,47] are shown in Figure 1(c) and 1(d), respectively.

WILEY-VCHIn the following, we focus mostly on the room temperature characterization of the transistor formed between leads 2 and 3 (see Figure 1(a)), with channel length $L$= 3.9 µm and average width $W$ =29.1 µm. Combined with leads 1 and 4, this transistor is an obvious choice for four-probe measurements. The device is intentionally measured after several months from the fabrication to study how the device performance is affected by the degradation due to air exposure of both the channel and the contacts. The Ti contact resistance, in particular, can be increased by the formation of $TiO_2$ due to oxygen diffusion under the metal leads, possibly facilitated by defects such as unreacted $MoO_3$. Indeed, devices on the same chip, measured soon after fabrication showed ohmic behavior with much lower contact resistance.[48]

**2.2. Diode characterization**

**Figure 2**(a) shows the current-voltage ($I_{ds} - V_{ds}$) characteristics between leads 2 and 3, used as the anode and the cathode of a two-terminal structure or, equivalently, as the drain and the source of the back-gated field-effect transistor formed with the substrate. The substrate is kept grounded ($V_{gs}$= 0 V), and the measurements are performed both in dark and under illumination by a white-LED at 5 mWcm$^{-2}$ light intensity. The dark curve shows a clear rectifying behavior with an exponentially growing current at negative voltages and a slower-growing current at positive voltages, reminiscent of a Schottky barrier diode on a p-type semiconductor. The current drastically increases upon illumination at low $|V_{ds}|$ bias and becomes quasi-saturated at $V_{ds}$ >10 V, such that the device can be proposed as an efficient photodetector. Indeed, it exhibits an excellent responsivity $R = I_{ph}/I_{opt}$= 2.5 AW$^{-1}$ at $V_{ds}$= 20 V ($I_{ph}$ is the photocurrent and $P_{opt}$ is the incident optical power), which slightly increases at higher biases.

For a defect-free Ti/MoS$_2$ interface, the barrier height $\Phi_B = \Phi_{Ti} - \chi_{MoS_2}$, resulting from the ideal Schottky-Mott relation,[44] would be null or negative considering the Ti work function $\Phi_{Ti}$= 4.33 eV [49] and the bilayer MoS$_2$ electron affinity $\chi_{MoS_2}$= 4.4 eV.[50] In general, metals with work function similar to the electron affinity of MoS2 are chosen as electrodes to achieve low Schottky barrier and therefore low contact resistance in field effect transistors. However, intrinsic defects, unavoidably present in MoS$_2$, are the cause of Fermi level pinning and originate a low Schottky barrier independently of the metal contact work function.[51] Indeed, rectifying Ti/MoS$_2$ bilayer contacts have been reported in previous studies.[52] Moreover, the Ti contacts are realistically characterized by a thin TiO$_2$ layers formed at the interface as consequence of the Ti deposition process performed at pressure (10$^{-8}$ mbar) above the ultrahigh vacuum condition (< 10$^{-9}$ mbar) needed to achieve Ti rather than TiO$_2$ at the MoS$_2$ interface.[53] Furthermore, the long exposure to air before measurement is expected to contribute to the oxidation process by diffusion. The presence of unreacted residuals of MoO3, which are visible on the flake as brighter dot branched sequences (Figure 1), likely facilitates the local Ti oxidation process. Consequently, the TiO$_2$ work function, $\Phi_{TiO_2}$= 4.9 – 5.5 eV,[54] reasonably explains the formation of a Schottky barrier either for electrons or holes. In addition, MoO3 can contribute to the formation of a Schottky barrier, being a high work function conducting material.[55]

Given the diode-like $I-V$ characteristic, as first option, we consider that leads 2 and 3 form a p-type Schottky junction and a more ohmic interface on the MoS$_2$ channel, respectively. In such case, the forward current can be qualitatively reproduced, for low injection at $V_{ds} < R_s I_{ds}$, by a simple Schottky diode Equation based on thermionic emission theory, corrected by a series resistance $R_s$ and an ideality factor $n$ to take into account possible deviations from the pure thermionic regime:[56]

$$I_{ds} = I_s \left( e^{\frac{q(V_{ds}-R_sI_{ds})}{nkT}} - 1 \right) \qquad (1)$$

q is the electron charge, $T$ is the temperature, $k$ the Boltzmann constant and $I_s$ is the reverse saturation current, which is constant in Equation (1). Here, $I_s$ can include a voltage dependence in more advanced theories, such as the thermionic-diffusion or the drift-current-limited models, which generally imply a dependence as the square root of the applied voltage.[57]

The fitting of Equation (1), as shown in the Figure 2(a), yields $R_s = 3.5 \times 10^7 \, \Omega$, which indicates a quite resistive device, $I_s = 40$ pA, and $n = 48$. The anomalously high ideality factor as well as the fact that MoS$_2$ is an n-type semiconductor, as we will show later, suggests that the p-type diode model is unsatisfactory and a deeper analysis of the carrier transport behavior is required.

We then consider a more realistic situation in which both leads form a Schottky junction on n-type MoS$_2$. Figure 2(b) compares the I-V curves obtained by swapping the role (as drain or anode) of leads 2 and 3. The curves are rectifying in both cases, but with increased forward (at $V_{ds} \leq 0$ V) and reverse (at $V_{ds} \geq 0$ V) currents, when using lead 3 as drain. Such difference cannot be ascribed to the slightly different geometrical area of the contacts. Indeed, this finding confirms that both contacts (2 and 3) are of Schottky type and present a little asymmetry, with possible different barrier heights, which we call $\phi_B^{(2)}$ and $\phi_B^{(3)}$, respectively. The two Schottky contacts constitute two back-to-back junctions in the lead-2/MoS$_2$ channel/lead-3 structure (the l2/MoS$_2$/l3 device, in short), as shown in Figure 2(c). The bias of any polarity of the l2/MoS$_2$/l3 device sets one Schottky barrier in the reverse direction and the other one in the forward direction. Hence, the current through the device is always limited by a reverse biased junction. This also explains the high series resistance. In particular, the current at $V_{ds} \leq 0$ V is limited by the reverse polarized junction 2 when lead 2 is used as the drain, or vice-versa by junction 3 when the drain is on lead 3. The lower value of the current when the drain is on lead 2 indicates that $\phi_B^{(2)} > \phi_B^{(3)}$. We notice that the rapid increase of the photocurrent at low bias, which can be observed in Figure 2(a) both at positive and negative voltages, is due to the expansion of the depletion region of either one of the reverse biased junctions, which improves the internal quantum efficiency of the l2/MoS$_2$/l3 device.

Despite the presence of a reverse biased junction, we are going to show that the combination of the two Schottky contacts can explain the observed rectifying behavior, with the higher (forward)/lower (reverse) current at negative/positive voltages, respectively.

Figure 2(d) shows that the logarithm of the l2/MoS$_2$/l3 device current, both in forward and reverse direction, is a linear function of the $\sqrt[4]{V_{ds} - R_s I_{ds}}$. A current dependence on the exponential of the fourth root of the applied voltage is typical of a reverse biased Schottky junction subjected to image force barrier lowering.[58-60] This suggests that the seemingly forward current of the l2/MoS$_2$/l3 device is actually the reverse current of the Schottky junction at contact 2, which increases exponentially because of the image force lowering of the barrier $\phi_B^{(2)}$. Indeed, such reverse current, which is voltage dependent, can be expressed as:[60]

$$I_s = AA^* T^2 \exp\left[\frac{q}{kT}\left(\alpha\sqrt[4]{|V_{ds} - R_s I_{ds}|} - \phi_B\right)\right] \tag{2}$$

Where A is the area of the junction, A* the effective Richardson constant, α a dimensional constant, and $\phi_B$ the Schottky barrier height at zero bias. According to Equation (2), the fitting of a straight line to the plot of $\ln(I_s)$ vs $\sqrt[4]{|V_{ds} - R_s I_{ds}|}$, as displayed in Figure 2(d), can be used to extrapolate the reverse saturation current $I_{s0}$ to zero bias ($\sqrt[4]{|V_{ds} - R_s I_{ds}|}= 0$ ). Such procedure has been iterated on the $I-V$ curves at different temperatures, displayed in Figure 2(e), to obtain $I_{s0}(T)$ and construct the Richardson plot of $\ln(I_{s0}/T^2)$ vs $1/T$, which according to Equation (2), and as confirmed in Figure 2(f), has a linear behavior,

$$\ln(I_{s0}/T^2) = \ln(AA^*) - \frac{q\phi_B}{kT} \tag{3}$$

$I_{s0}$ can be extrapolated with the same procedure from the reverse current of the l2/MoS$_2$/l3 device of Figure 2(e) as well. In such case, the reverse biased junction at contact 3, i.e. the barrier $\phi_B^{(3)}$, limits the device current. The Richardson plots from the zero-bias extrapolated forward and reverse currents of the l2/MoS$_2$/l3 device (Figure 2(f)) and inset of Figure 2(f), respectively) are used to estimate the effective Richardson constant and the zero-bias Schottky barrier heights from the intercept and the slope of the linear fitting. We obtain $A^* = 8 \times 10^{-6} \text{Acm}^{-2}\text{K}^{-2}$, $\phi_B^{(2)} = 0.51$ eV and $\phi_B^{(3)} = 0.31$ eV. The lower value of $\phi_B^{(3)}$ confirms the expected asymmetry of the two barriers, which can be due to a different oxidation or defect density at the two contacts. The unreacted residuals of $MoO_3$, which are randomly distributed, could greatly contribute to the asymmetry.

The measured effective Richardson constant is several orders of magnitude lower than what has been measured using few-layer $MoS_2$/graphene heterojunctions or estimated by theoretical calculations.[43] The discrepancy is likely due to the oxide layer at the contacts, which limits the effective injection area to a narrow central region of the geometrical area, where the $TiO_2$ layer might be thinner, or because it introduces a tunneling factor.[56,60] In the latter case, the measured Richardson constant is the product of the effective Richardson constant and a tunneling factor through the oxide layer: $A^*_{Meas.} = A^* exp(-\chi^{1/2}\delta)$, where $\chi[\text{eV}]$ is the mean barrier height and $\delta[\text{Å}]$ is the thickness of the oxide.

In Figure 2(f) we also show that $R_s$ decreases for rising temperature as expected for a semiconductor, indicating that the series resistance is dominated by the $MoS_2$ channel.

The above considerations lead to the qualitative energy band model shown in **Figure 3**, which fully explains the observed $I - V$ curves. The key points are: *(i)* The barriers are slightly asymmetric. *(ii)* The gating effect of the grounded substrate shifts the band diagram closer to the junction connected to the forcing lead (the anode or drain) downwards for negative $V_{ds}$ (thus reducing the Schottky barrier) and upwards for positive $V_{ds}$ (thus increasing the Schottky barrier). *(iii)* The applied bias drops mainly on the junction connected to the forcing lead, as proved by Kelvin force microscopy [61] and confirmed by four-probe measurements in this study. This behavior is obvious when the forced junction is reverse biased, as it occurs for $V_{ds} < 0$ V, and is caused by the gating effect, which empties the channel from free carriers close to the drain region, when the powered junction is forward biased, i.e. for $V_{ds} > 0$ V. *(iv)* The image force barrier lowering affects the current mainly when the forcing electrode is on the reverse biased junction thereby causing the high forward current at $V_{ds} \leq 0$ V. *(v)* When the grounded lead is on the reversed biased junction, the barrier is lightly affected by the image force barrier lowering. In such case, the low field at the grounded junction, which is about in a flat-band condition, strongly suppresses the current. *(vi)* Due to the high barrier ($\geq 1$ eV), the contribution of the holes, which are the minority carriers, can be neglected.

**2.3. Transistor characterization**

The source-drain $I_{ds} - V_{ds}$ output characteristics of the transistor as a function of the gate voltage, in dark, are reported in **Figure 4**(a). An increase of the current is observed for increasing $V_{gs}$, which makes the channel more conductive by shifting the energy bands downwards, as sketched in Figure 2(c). The gate voltage does not change the general rectifying behavior, although the rectification ratio, here defined as the current ratio at $\mp 15$ V, increases at negative $V_{gs}$, as shown in Figure 4(b). This behavior can be easily understood considering that the negative gate voltage, which shifts the channel bands upwards (Figure 2(c)), affects mainly the reverse current of the l2/MoS$_2$/l3 device. Referring to Figure 3, it can be easily understood that the up-shift flattens the bands and increases the effective Schottky barrier at the source, while enhances the band bending at the drain, where for $V_{ds} > 0$ V a high drain-to-gate $V_{dg} = V_{ds} - V_{gs}$ voltage drop occurs. Flatter bands mean reduced electron injection from the source and suppressed reverse current or increased rectification ratio. The effect of the negative $V_{gs}$ is less pronounced on the forward current of the device since the injecting junction (i.e. the negatively biased drain) is in this case subjected to a lower $V_{dg}$ voltage drop. The increase of the rectification ratio at $V_{gs} < 0$ V is less evident under illumination (Figure 4(b)) since the photogeneration strongly increases the l2/MoS$_2$/l3 device reverse current.

The transfer characteristics $G_{ds}$ versus $V_{gs}$ curves, where $G_{ds} = I_{ds}/V_{ds}$ is the conductance between source and drain of the l2/MoS$_2$/l3 device, are shown in Figure 4(c) and 4(d) for two different drain biases: $V_{ds} = \pm 20V$. These curves reveal a normally-on n-type device with threshold voltage $V_{th} \sim -25$ V for the forward sweep. The n-type behavior of few-layer MoS$_2$ has been reported and discussed in several studies.[62-67] It is generally attributed to dissociative adsorption of oxygen on the surface defects of MoS$_2$ that locally lowers the conduction band edge favoring n-doping.[68]

Figure 4(c) and 4(d) also show that light enhances the channel conductivity and reduces the threshold voltage. The left shift of the threshold voltage, that lowers the on/off ratio over the measured $V_{gs}$ range, is the so-called photo-gating effect and has been discussed in previous works.[66,69-72] It is caused by photocharge trapping in intrinsic trap centers such as vacancy defects, substitutional defects, and dislocation cores at the grain boundaries of MoS$_2$ or in extrinsic defects due to O$_2$, H$_2$O or other process residues adsorbed on the exposed MoS$_2$ surface or at the MoS$_2$/SiO$_2$ interface. Both the trapping and detrapping processes can be very slow and cause a persistent photoconductivity, as demonstrated in Figure 4(e), displaying the transient behavior of the current in a dark-light-dark sequence. Both the raising and decay current can be fitted by a double exponential, with a faster ($\sim 120$ s) and a longer ($> 1000$ s) time constant, corresponding to shallower and deeper traps, respectively. We highlight that the device remains in a higher conductive state (persistent photoconductivity) even 4000 s after the light switching off.

While the persistent photoconductivity has been attributed to the capture of photogenerated holes in deep inter-gap states,[66] the faster component observed in the photocurrent is indicative of a high density of shallower trap states. Such states are detrimental to the channel carrier mobility. Their effect can be observed in dark from the behavior of the channel current when the l2/MoS$_2$/l3 device is switched between the on and off state by a gate voltage pulse as shown in Figure 4(f). Following a short switching-on pulse, the drain current returns to the off state in a few seconds (the time needed to empty the channel from the majority carriers) and after that exhibits a slower decay due to the detrapping of electrons from the shallower traps, with a consistent time constant around 100 s.

The same intrinsic and extrinsic trapping mechanism has been identified also as the cause of the wide hysteresis observed in Figure 4(d) between the reverse and forward gate voltage sweeps.[67,73-77]

In the attempt to separate the intrinsic properties of the MoS$_2$ channel from the effect of the rectifying contacts, we repeated the characterization using a four-probe method, with the forcing current between leads 1 and 4 and the sensing voltage between leads 2 and 3. **Figure 5**(a) and (b) show the $I-V$ characteristics. The $V_{ds}$ range between contacts 2 and 3 is asymmetric and is set by the $\pm 20$ V limit that we allowed at the current-forcing electrode 1. As expected, the output characteristics are more symmetric. Remarkably, the output $I_{ds}-V_{ds}$ curves show an almost linear behavior for $V_{gs} \geq -10$ V, corresponding to the "on" state of the transistor (Figure 5(c)), while they scale as a power of $V_{ds}$ at more negative gate voltages (Figure 5(d)), that is when the transistor approaches the "off" state. The same behaviour can be observed in Figure 5(a) comparing the higher current under illumination to the lower current in dark. Non-linear output characteristics have been often reported in MoS$_2$ transistors and attributed to the existence of an exponential distribution of trap states in the MoS$_2$ bandgap.[78] Our findings indicate that the previously mentioned high density of shallow traps cause a space charge limited conductance in the low injection regime, and that the same traps become less effective in modifying the conduction when there is abundance of free carriers in the channel. Figure 5(e) and 5(f) confirm the n-type conduction, the photoconductivity, and the hysteresis observed with two-probe measurements. This observation remarkably confirms that such features are related to the MoS$_2$ channel rather than being originated by the metal/MoS$_2$ interfaces.

The extra series resistances introduced by the reverse biased junctions in our measuring setup, as well as the higher trapping efficiency at low current, are the reason of the seemingly lower channel conductance measured by the four-probe method compared to two-probe setup. Indeed, the resulting bias between leads 2 and 3 is only $|V_{ds}| \leq 1$ V in four-probe configuration (Figure 5 (e)) while it is $|V_{ds}| \approx 20$ V in two-probe measurements (Figure 4 (c)).

As can be observed from Figure 4(c) or 5(e), in the "on" state, the transfer characteristic has a quadratic dependence on $V_{gs}$, despite the FET is in the triode region for which $V_{ds} < (V_{gs} - V_{th})$, where a linear $G_{ds} - V_{gs}$ dependence usually occurs in traditional semiconductor long-channel FETs. The quadratic dependence is quite common in MoS$_2$ transistors.[6, 22, 75, 79] To explain it, a model based on a gate voltage dependent mobility has been proposed.[67] Accordingly, we write the mobility as $\mu = \mu_V(V_{gs} - V_{th})$, and the $G_{gs} - V_{gs}$ relationship of the FET in the triode region becomes:

$$G_{ds} = \mu_V C_{ox} \frac{W}{L} (V_{gs} - V_{th})^2 \quad (4)$$

where $\mu_V$ is expressed in cm$^2$V$^{-2}$s$^{-1}$, $C_{ox} = \varepsilon_{SiO_2}/t_{SiO_2} = 6.9 \cdot 10^{-9}$ Fcm$^{-2}$ is the capacitance per unit area of the SiO$_2$ gate dielectric, $L$ and $W$ are the channel length and width, respectively, and $V_{th}$ is the threshold voltage.

A linear dependence of the mobility on $V_{gs}$ can be justified considering that a growing carrier density, induced by an increasing gate voltage, enhances the screening of Coulombic scattering potentials and results in higher mobility.[80] A mobility dependent on a power of the gate voltage ($\mu = \mu_V(V_{gs} - V_{th})^\alpha$ with $\alpha \geq 1$) has been suggested for organic and hydrogenated amorphous silicon (a-Si:H) FETs.[81] In materials with a high density of inter-gap states, as the MoS$_2$ bilayer under study, a growing gate voltage favors the filling of the traps, which become gradually inactive in limiting the mobility by continuous capture and emission processes.[82]

The fitting of Equation (4) results in a maximum mobility of ~0.01 cm$^2$V$^{-1}$s$^{-1}$ at $V_{gs} = 50$ V, which is below the range $0.1 - 100$ cm$^2$V$^{-1}$s$^{-1}$ typically reported for uncovered CVD MoS$_2$ transistors on $SiO_2$.[66,70,71,74,83] Obviously, process residuals, surface roughness and $MoO_3$ patterns (visible on the flake), which could hide grain boundaries, strongly affect the mobility,[84] together with the already mentioned high density of shallow traps due to intrinsic defects. Chemisorbed oxygen and water due to the long air exposure further suppresses the current and the mobility.[22,67,85]

## 3. Conclusion

We have presented the high-bias characterization of a back-gated transistor with bilayer $MoS_2$ channel and Ti rectifying contacts. We have proposed injection of electrons over the two barriers, one of which is effectively lowered by image force, as the main transport mechanism. We have shown that the device has a high photoresponsivity and a persistent photoconductivity. We have discussed the output and transfer characteristics of the FET, measured by both two and four-probe configuration, and argued that the observed hysteresis and persistent photoconductivity are channel proprieties rather that effects caused by the contacts.

## 4. Experimental Section

$MoS_2$ flakes were synthesized by chemical vapor deposition at $T = 750$ K over a highly doped p-Si substrate ($\rho \sim 0.01$ Ωcm) capped by a 500 nm $SiO_2$ layer. We used S and $MoO_3$ powders as reactants, in two ceramic boats placed next to each other inside a tube furnace. The Si/$SiO_2$ substrate was mounted on the top of the boats, faced down. The resulting $MoS_2$ flakes, with prevailing triangular and hexagonal shapes, were mostly bilayers or multilayers, although single layers sometimes occurred.

Figure 1(a) shows the scanning electron microscopy (SEM) image of a selected flake with metal contacts deposited by standard electron-beam lithography (EBL) and lift-off process. Unreacted residuals of $MoO_3$ are visible on the flake as brighter dot branched sequences. We evaporated 20 nm Ti (at $5 \times 10^{-8}$ mbar) as the contact layer and $130\ nm$ Au (at $5 \times 10^{-8}$ mbar) as the cover layer. The schematic cross-section of the device is shown in Figure 1(b). The Si substrate is used as the common back-gate of the FETs formed between any couple of contacts. The multi leads and the different spacing enable four-probe as well as TLM measurements to study the channel and the contact resistance as well as short channel effects in the transistor.

The MoS$_2$ flake was characterized by micro-Raman. A typical spectrum and the map, shown in Figure 1(c) and 1(d), confirm an average separation of $21-22$ cm$^{-1}$ between the $A_{1g}$ (out-of plane vibration of S atoms) and the $E_{2g}^1$ (in-plane vibration of Mo and S atoms) peaks. Such separation is typical of a bilayer MoS$_2$.[46,47] The bilayer nature of the flake was further confirmed by atomic force microscopy (AFM) which measured a step height of ~1.2 mn.

We performed electrical measurements in a probe station (Janis ST-500) connected to a Keithley 4200 semiconductor characterization system (SCS), at given temperatures (in the range $200-400$ K) and pressure (~$2-5$ mbar).

Most of the measurements were performed at room temperature on the transistor formed between leads 2 and 3 (refer to Figure 1(a)), with channel length $L = 3.9$ μm and average width $W = 29.1$ μm. Combined with leads 1 and 4, this transistor is an obvious choice for four-probe measurements. Other two-contacts combinations give similar or slightly lower current and lead to the same conclusions. The characterization was performed both in two and four-probe configuration. Drain and back-gate were biased at high voltage, up to $\pm 60$ V range, the limitation being the request of preventing dielectric breakdown or to keep the dielectric leakage current $< 1$ pA. We highlight that the device was measured after several months from the fabrication to study the effect of air-exposure which is expected to result in oxidation of the Ti due to oxygen diffusion under the contacts.

**Acknowledgements**

We thank Dr. E. Enrico and F. Pillepich (Istituto Nazionale di Ricerca Metrologica, INRIM, Italy) for their contribution to the device fabrication and to Raman characterization. We are grateful to Dr. Yee Sin Ang (Singapore University of Technology and Design, SUTD, Singapore) for the useful discussion on data analysis.

Received: ((will be filled in by the editorial staff))

Revised: ((will be filled in by the editorial staff))

Published online: ((will be filled in by the editorial staff))

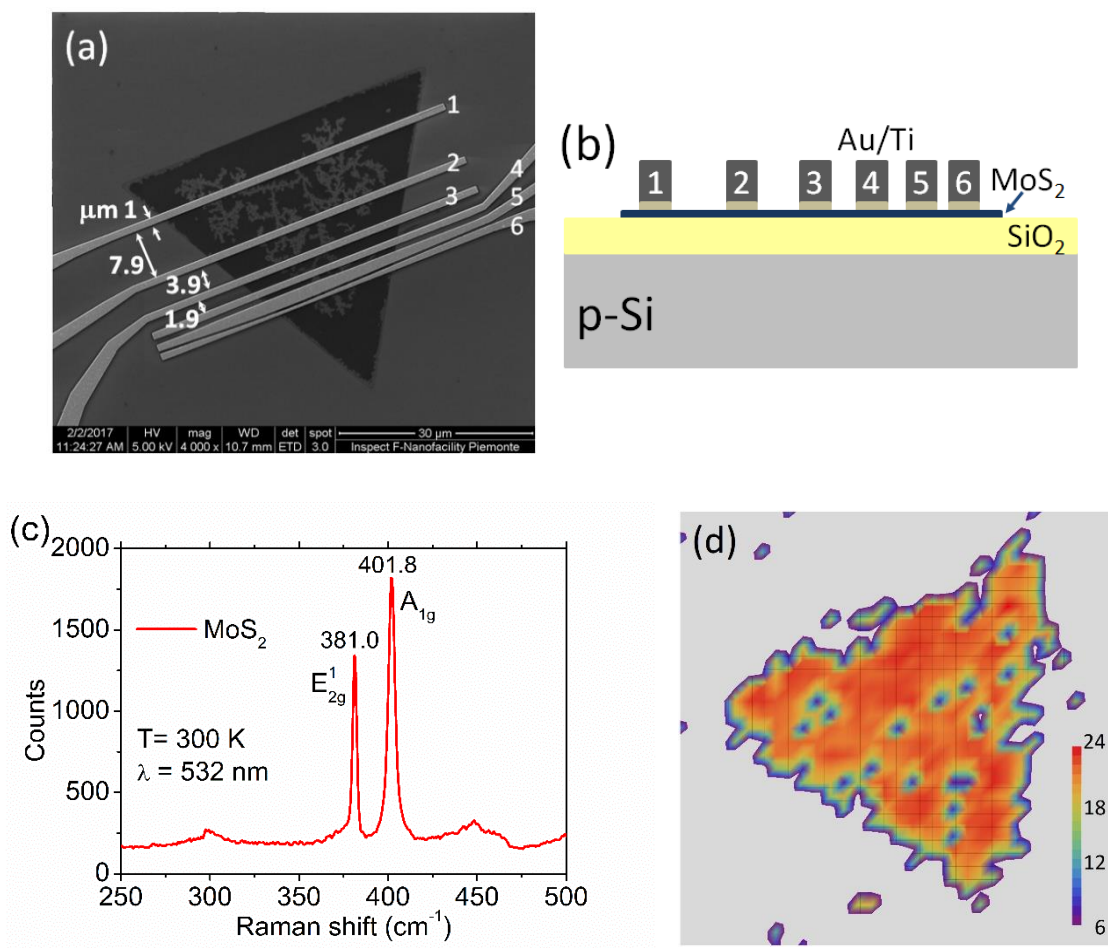

**Figure 1.** (a) SEM top view of a CVD-synthesized bilayer MoS$_2$ with Ti/Au contacts. (b) Schematic of the back-gate MoS$_2$ transistors (TLM structure). (c) Raman spectrum of the bilayer MoS$_2$. (d) Map of the difference between $A_{1g}$ and $E_{2g}^1$ peaks of micro-Raman spectra.

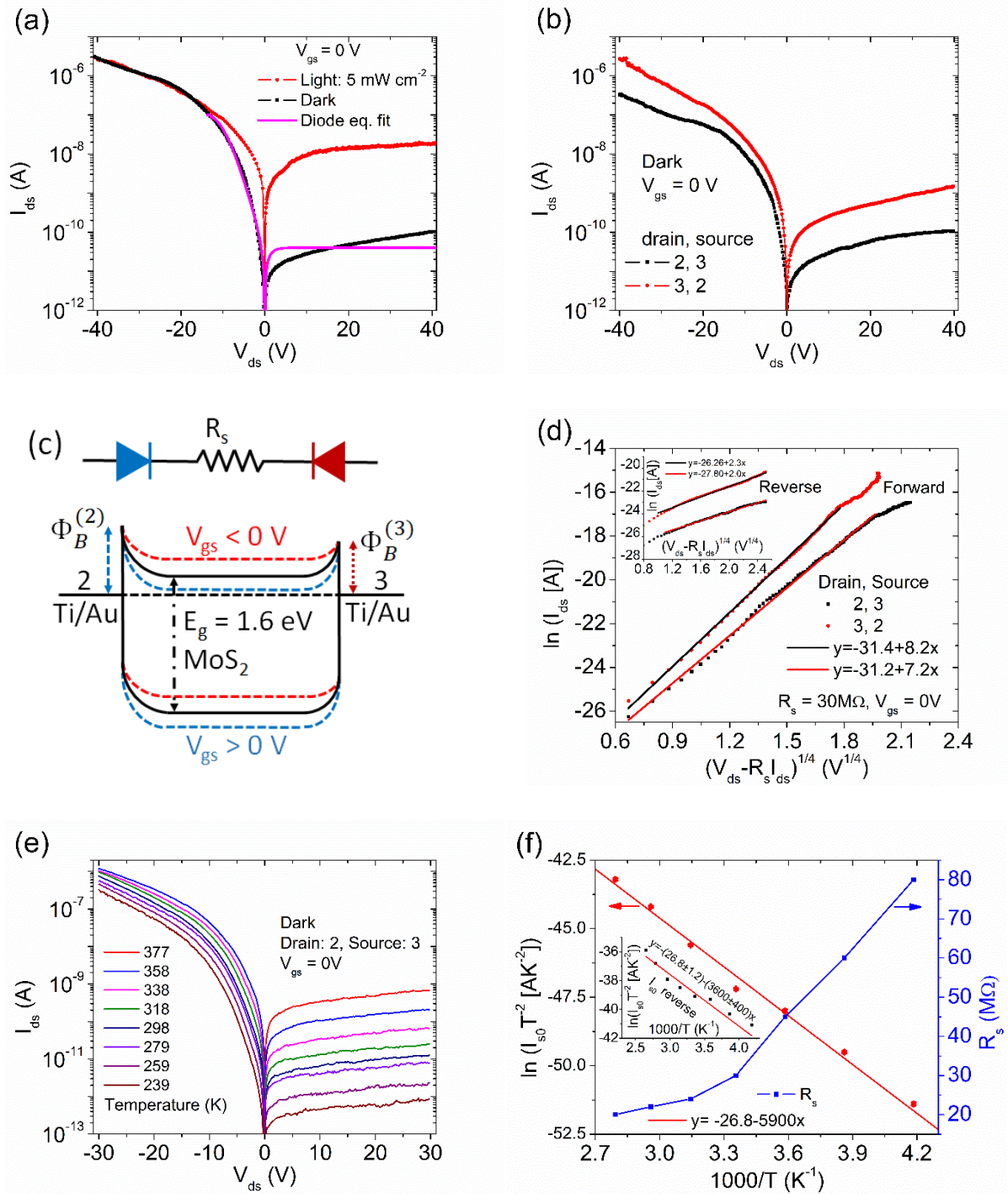

**Figure 2.** (a) Two-probe I-V characteristics in dark and under illumination (5 $mWcm^{-2}$ white-LED light) for the device of Figure 1, with contacts 2 and 3 used as the drain and the source (or the anode and the cathode), respectively. The gate is grounded. The solid line (magenta curve) is the fit of the Schottky diode Equation (1). (b) I-V characteristics in dark and at grounded gate ($V_{gs} = 0$) with swapped role of lead 2 and 3. For the two curves, leads 2 and 3 are used as the drain in turn. (c) Schematic of the device consisting of two back to back diodes and a series resistance, and band-diagram with two asymmetric junctions with barrier heights $\phi_B^{(2)}$ and $\phi_B^{(3)}$, respectively, and effect of the gate voltage. (d) Semi-logarithmic plot of drain-source dark current, $I_{ds}$, for $V_{ds} \leq 0\ V$ (forward current of the l2/MoS2/l3 device) as a function of the fourth root of the voltage drop across the junction, $\sqrt[4]{V_{ds} - R_s I_{ds}}$, and fit of Equation (2). The inset shows the same plot for $V_{ds} \geq 0\ V$ (reverse current of the l2/MoS2/l3 device). (e) I-V characteristics at different temperatures between contacts 2 and 3 used as the drain and the source, respectively. (f) Richardson plot and temperature behavior of the series resistance $R_s$ obtained from the zero-bias current extrapolated from the forward or the reverse (inset) $I_{ds}$ current.

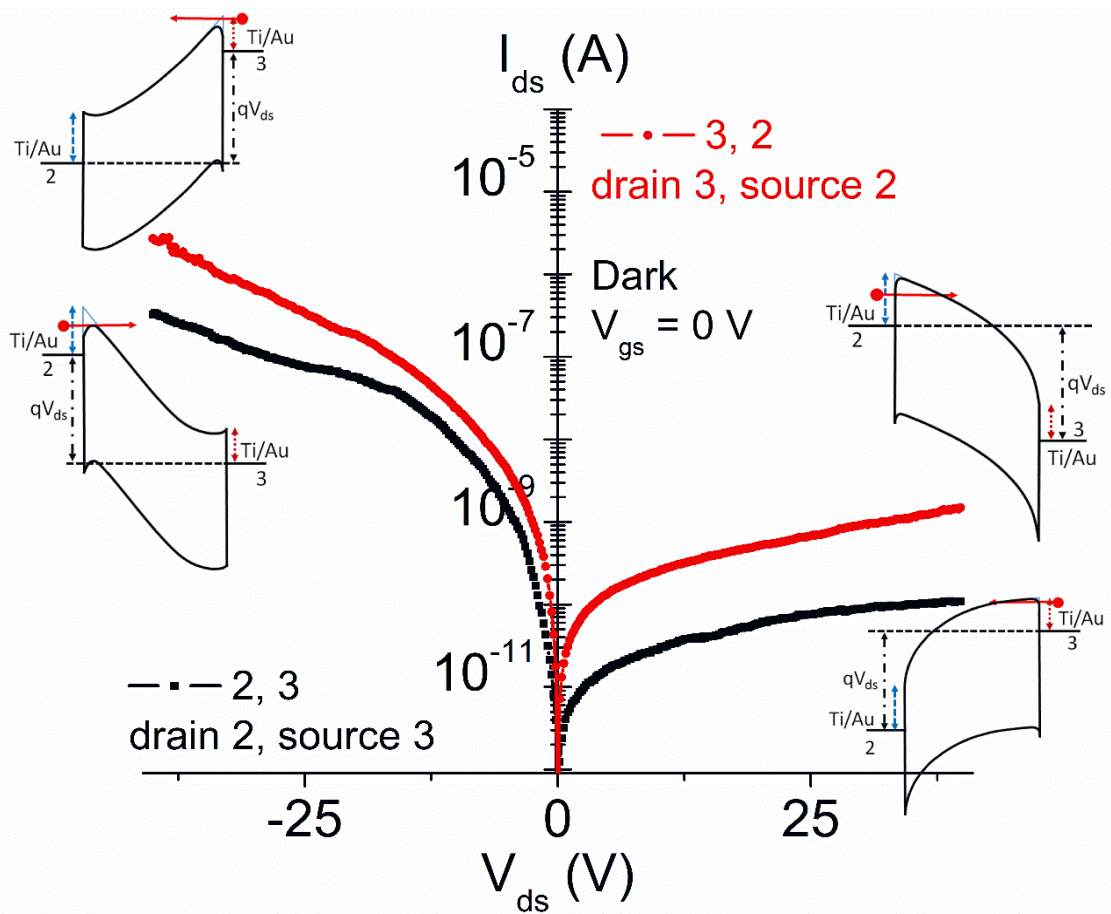

**Figure 3.** Band diagram based on two back-to-back Schottky barriers. The forward current observed for negative $V_{ds}$ is due to the image force barrier lowering at the forced junction, while the lower (reverse) current at $V_{ds} > 0\ V$ is limited by the low electric field at the grounded junction. The red arrow represents the direction of the electron flow.

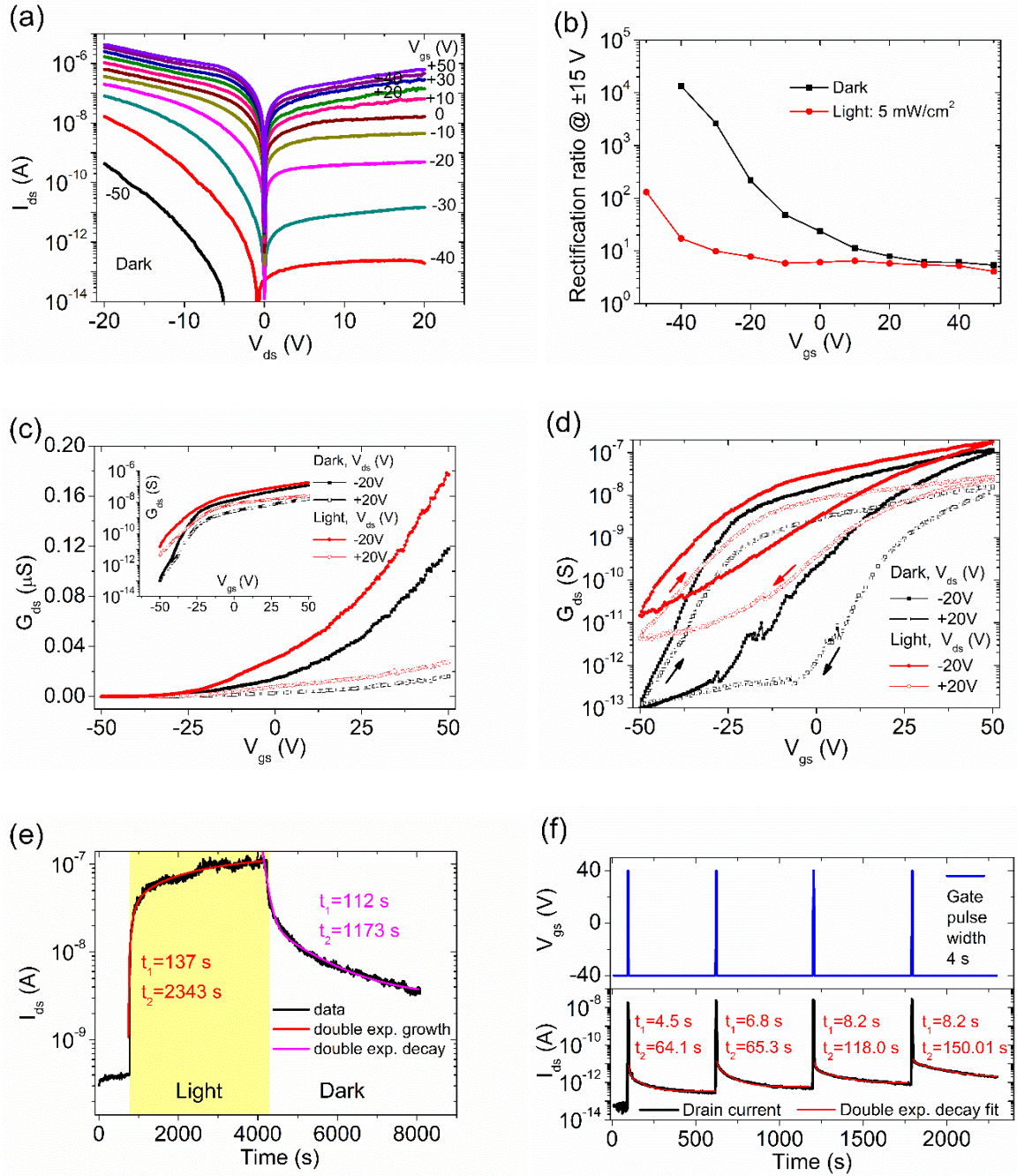

**Figure 4.** Two-probe transistor characterization, with contact 2 and 3 used as the drain and the source, respectively. Voltages are referred to the source, which is grounded. (a) Output $I_{ds} - V_{ds}$ characteristics at different $V_{gs}$ in dark. (b) Rectification ratio at $V_{ds} = \mp 15\ V$ in dark and under illumination ($5\ mW cm^{-2}$ white-LED light) as a function of $V_{gs}$. High-bias ($V_{ds} = \pm 20\ V$) transfer characteristics for forward (c) and forward-backward (d) sweeps of the gate voltage, in dark and under illumination (the inset in (c) shows the conductance in logarithmic scale). (e) Transient behavior of the l2/MoS2/l3 current in a dark-light-dark cycle at $V_{ds} =$

$-20\ V$ and $V_{gs} = 0\ V$. (f) Transient behavior of the source-drain current at $V_{gs} = -40\ V$ ("off" state) upon application of short $V_{gs} = 40\ V$ pulses, in dark.

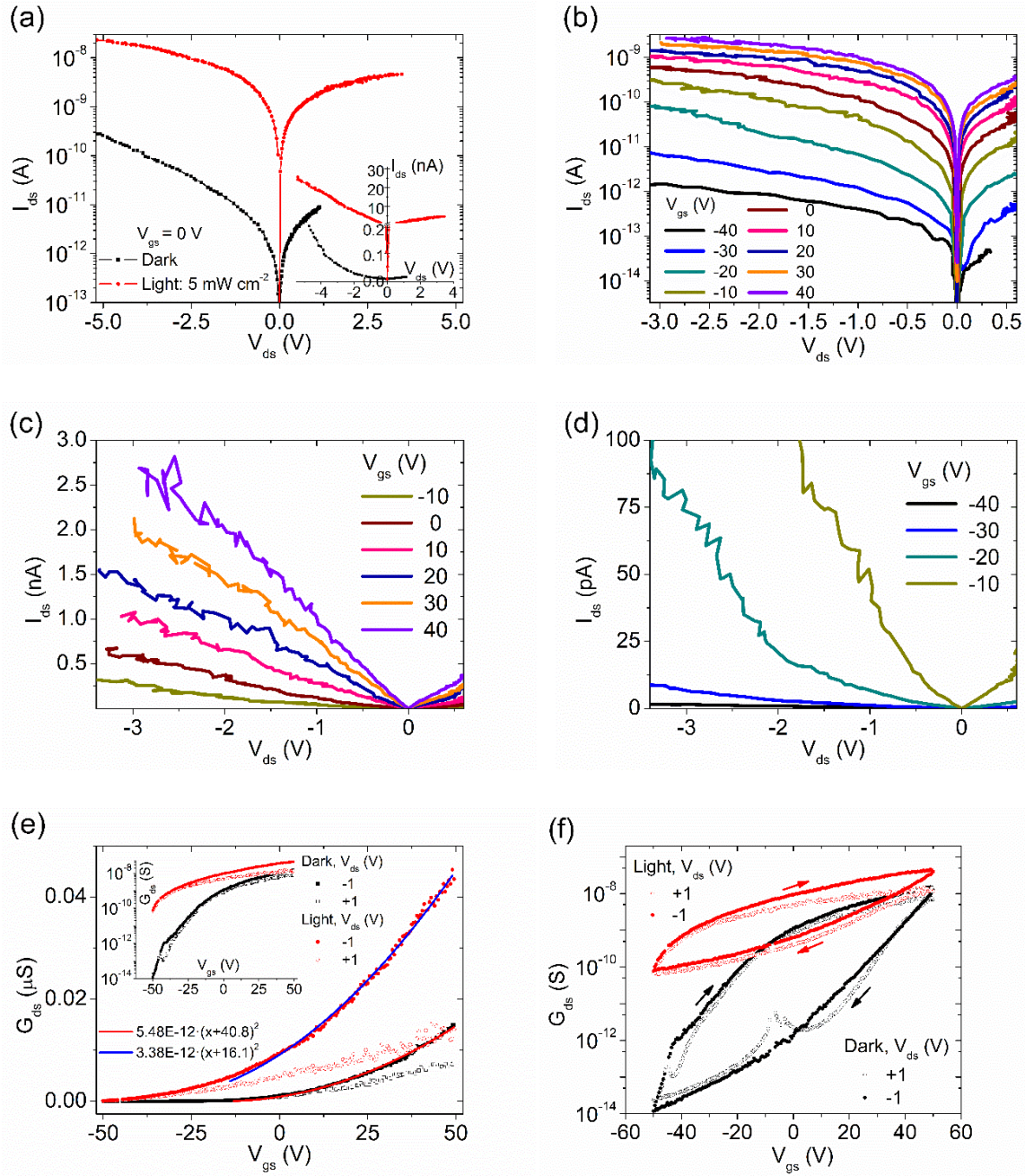

**Figure 5.** Four-probe electrical characterization with the forcing current between leads 1 and 4 and the sensing voltage between leads 2 and 3. (a) Output I-V characteristics at grounded gate ($V_{gs} = 0\ V$), in dark and under $5\ mWcm^{-2}$ white-LED light. Output characteristics in dark at different gate voltages with source-drain current in logarithmic scale (b) and in linear scale, (c) and (d). Transfer characteristics in dark and under illumination at $V_{ds} = \pm 1\ V$ for a forward $V_{gs}$

sweep (e) and for a complete forward-reverse sweep (f). Inset of plot (e) shows the current in logarithmic scale.

**Table of contents**

**Ti Schottky contacts on molybdenum disulfide (MoS$_2$) result in rectifying current-voltage output characteristics, which can be explained by image-force barrier lowering at metal/MoS$_2$ interfaces.** The two back-to-back Schottky junctions can be exploited for efficient photodetection. Transistor features such as hysteresis and persistent photoconductivity, which are due to charge trapping in intrinsic and extrinsic defects, are properties of the MoS$_2$ channel rather than effects of the contacts.

**Keyword**

**Schottky MoS$_2$ transistor**

A. Di Bartolomeo*, A. Grillo, F. Urban, L. Iemmo, F. Giubileo, G. Luongo, G. Amato, L. Croin, L. Sun, S.-J. Liang, and L. K. Ang

**Title**

**Asymmetric Schottky contacts in bilayer MoS$_2$ field effect transistors**

ToC figure

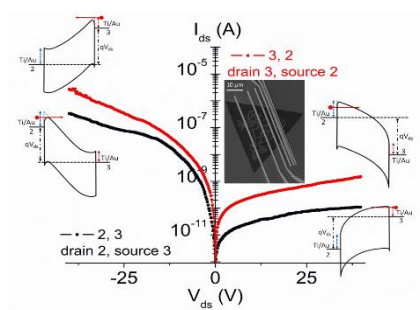